\documentclass[twocolumn,english,aps,pra,nofootinbib,superscriptaddress,showpacs,showkeys,longbibliography]{revtex4-1}
\usepackage[T1]{fontenc}
\usepackage[latin9]{inputenc}
\usepackage{color}
\usepackage{babel}
\usepackage{bbold}
\usepackage{amssymb}
\usepackage{amsmath}
\usepackage{mathtools}
\usepackage{siunitx}
\usepackage{verbatim}
\usepackage{graphicx}
\usepackage{subfigure}
\graphicspath{ {images/} }
\usepackage{bbm, dsfont}
\usepackage{braket}
\usepackage[colorlinks=true,
    linkcolor=red, 
    citecolor=blue, 
    urlcolor =blue]{hyperref}
\makeatletter
\usepackage{amsfonts}
\usepackage{babel}
\usepackage{braket}
\newcommand\minus{
  \setbox0=\hbox{-}
  \vcenter{
    \hrule width\wd0 height \the\fontdimen8\textfont3
  }%
}

\makeatother

\begin{document}

\title{Encrypt me! A game-based approach to Bell inequalities and quantum cryptography}
\author{Andrea L{\'o}pez-Incera}
\affiliation{Institute for Theoretical Physics, University of Innsbruck, A-6020 Innsbruck, Austria}
\email{andrea.lopez-incera@uibk.ac.at}
\author{Andreas Hartmann}
\affiliation{Institute for Theoretical Physics, University of Innsbruck, A-6020 Innsbruck, Austria}
\email{andreas.hartmann@uibk.ac.at}
\author{Wolfgang D{\"u}r}
\affiliation{Institute for Theoretical Physics, University of Innsbruck, A-6020 Innsbruck, Austria}
\email{Wolfgang.Duer@uibk.ac.at}
\begin{abstract}
We present a game-based approach to teach Bell inequalities and quantum cryptography at high school. The approach is based on kinesthetic activities and allows students to experience and discover quantum features and their applications first-hand. We represent quantum states by the orientation of students, and mimic quantitative random behaviour and measurements using dice and apps.
\end{abstract}
\pacs{}
\maketitle

\section{Introduction}
Quantum technologies are at the brink to be commercially utilized, and we are entering an era where quantum mechanics is no longer just a topic that is relevant for scientists who study puzzling effects in their notebooks or labs, but also for engineers or even the general public. Quantum random number generators and quantum cryptography systems are already commercially available \cite{RevModPhys.89.015004, PhysRevX.4.031056, RevModPhys.74.145}, and researchers and some companies have started to build prototypes of quantum computers or quantum simulators \cite{Ac_n_2018, alvarez2018quantum, Behera2019, li2017approximate, boixo2014quantum, 45919, 44815, 46227, 45467}, and to investigate their usage to solve relevant problems in science, economy and beyond. Quantum networks, or as some people call it a quantum internet \cite{quantuminternetWolfgang, wehner2018quantum, kimble2008quantum}, might well be a reality within the near or mid future, and may have a similar influence as the development of the classical internet which plays an important role in society and everyone's daily life. More technical applications of quantum technologies, e.g. for high precision measurements (quantum metrology) are also discussed, and there is a huge effort worldwide to further develop such quantum technologies as is expressed in flagship programs in various countries or areas.

In order to further develop and use such technologies, the best minds are needed. This makes it an important goal to include such topics in the standard curriculum at high school, and to develop ways to teach them in a modern and interesting way. Raising the awareness of the potential of such quantum technologies, together with a basic understanding of their functionality, can also be considered a goal for general education. This paper aims at providing a treatment of the most advanced and immediate application of quantum technologies, namely quantum cryptography. At the same time, we treat a topic of fundamental interest that even goes beyond quantum mechanics, namely Bell inequalities \cite{RevModPhys.86.419, Bell1, PhysRevLett.115.250402, PhysRevLett.115.250401}. This is a subject that highlights the puzzling and counter-intuitive nature of our world, and contributes to the general theme of teaching on nature of science. While this is usually not part of a standard physics curriculum, we nevertheless believe it is of importance and relevance, and shows what natural sciences can tell us about the world we live in.

In this work, we take a game based approach to these topics \cite{doi:10.1119/1.5036620,goff,kohnle,scienceathome}, where we follow Ref.~\cite{LopezDur2019}. There, a game was introduced that allows students to take on the role of quantum bits (qubits) and scientists, and discover the features and rules of the quantum world. The basic principles of quantum mechanics including the superposition of states, the behavior under measurements as well as entanglement can be treated together with advanced topics such as decoherence. A key ingredient are kinesthetic activities \cite{McSha00, Begel:2004:KLC:1028174.971367, Sin10, Bru16} that allow students to directly experience these features, which supports a better understanding and helps assimilate concepts. One nice feature is that students are supposed to discover underlying rules from experimental data, thereby taking on the role of true scientists. Here we adapt and extend this setting to treat slightly more complex cases, including entanglement-based quantum cryptography \cite{PhysRevLett.84.4729, ursin2007entanglement, PhysRevA.63.012309} and Bell inequalities. While it suffices to consider only two different bases ($z$ and $x$) and hence four different states to understand basic features (and also quantum cryptography based on the BB84 protocol \cite{Bennett_2014}), more states and settings are required here. We hence slightly adopt the representation of quantum states and measurements, identifying orientation in space with direction in the Bloch circle (a reduced version of Bloch sphere where complex coefficients are avoided). The behavior of the quantum qubits is probabilistic, and depends on the relative orientation between the measurement axis (corresponding to the measurement basis) and the quantum state, i.e., the orientation of the qubit (or student) in space. In order to properly mimic the required probability distributions we suggest the usage of dice (either standard, tetrahedron or multiple ones). This can be done using real dice or freely available apps. Together, this allows one not only to understand the underlying principles of quantum cryptography, but also to experience the puzzling features of quantum mechanics in a qualitative and quantitative way first-hand.

The paper is organized as follows. In Section \ref{Main proposal} we describe our main approach. We introduce the setting and representation we use for one qubit and entangled pairs of qubits. In Section~\ref{bell test}, we provide background on Bell inequalities (CHSH \cite{PhysRevLett.23.880} and a variant by Mermin \cite{Mermin} that allows for a simplified treatment) and how we can include Bell tests in the framework of our game. In Section \ref{QuantumCryptography} we consider the application to quantum cryptography. Each of the mentioned sections is divided into the \textit{theoretical background} section, which is needed to understand the physics behind our proposal, and the \textit{rules of the game} section, where we specify the game and its rules, and how to implement it in class. We also provide an estimate of the required statistics to observe the desired effects (Appendix~\ref{app statist}). This is necessary as we are interested in expectation values of random processes with intrinsic fluctuations.

\section{Main Proposal}\label{Main proposal}
We propose a game to illustrate advanced concepts of quantum mechanics such as entanglement, Bell tests or quantum cryptography. The students play the role of both qubits and scientists and the measurement results that they obtain as scientists are identical to the ones they would obtain in a real laboratory. Therefore, the students can experience first-hand the puzzling features of quantum mechanics and have to come up with conclusions and explanations for the results they obtain.

In an initial work \cite{LopezDur2019}, a game is proposed in which students play the role of both, scientists and qubits, in such a way that quantum states are reproduced by different positions of arms and legs and quantum measurements by body movements. The possible measurement processes that can be mimicked within this framework are measurements along the $x$ and $z$ directions.

In this work, we present an alternative and complementary approach that allows the scientists to measure the qubits in more directions than just the orthogonal $x$ and $z$. Thus, more advanced concepts and experiments such as the Bell test can be introduced without the need of a strong mathematical background.

In this approach, the class is split into two groups: one plays the role of qubits and the other one the role of scientists. The goal of the scientist-students is to prepare qubit-students in certain states and measure them by hitting them with a ball. The qubit-students try to avoid the balls and they have to follow certain rules of the game (see Sections~\ref{one qubit} and \ref{entangled qubits} for  details) in order to mimic the real qubits, including both the stochastic behaviour and the state change after the measurement. Afterwards, the scientist-students analyze the measurement results, make conclusions and find theories that explain what has been observed.

The game is thus designed such that the students can learn not only the theoretical features of quantum mechanics, but also experience how to be a real scientist and what this implies. In particular, the analysis of the results and the critical thinking that should be developed in order to come up with suitable explanations are of high importance in our approach. Dealing with the puzzling features of quantum mechanics gives also an opportunity to enhance critical analysis since the experimental results do not match the previous, classical intuition and knowledge that the students may have in advance.

In the following sections, we explain the rules for the students in order to behave like actual qubits. The scientists can prepare either single-qubit states or a pair of qubits in an entangled state. We thus describe each case separately. Depending on the activity, the teacher can choose to work with only single-qubit states or with entangled states. For instance, only single-qubit states are needed to play the BB84 cryptography protocol (Section~\ref{BB84}), but one needs entangled qubits to play the Bell test (Section~\ref{bell test}) or the E91 protocol (Section~\ref{E91}).

\subsection{Single qubit}\label{one qubit}
\subsubsection{Theoretical background}\label{theo one qubit}
\begin{figure}[ht!]
\includegraphics[width=3.4in]{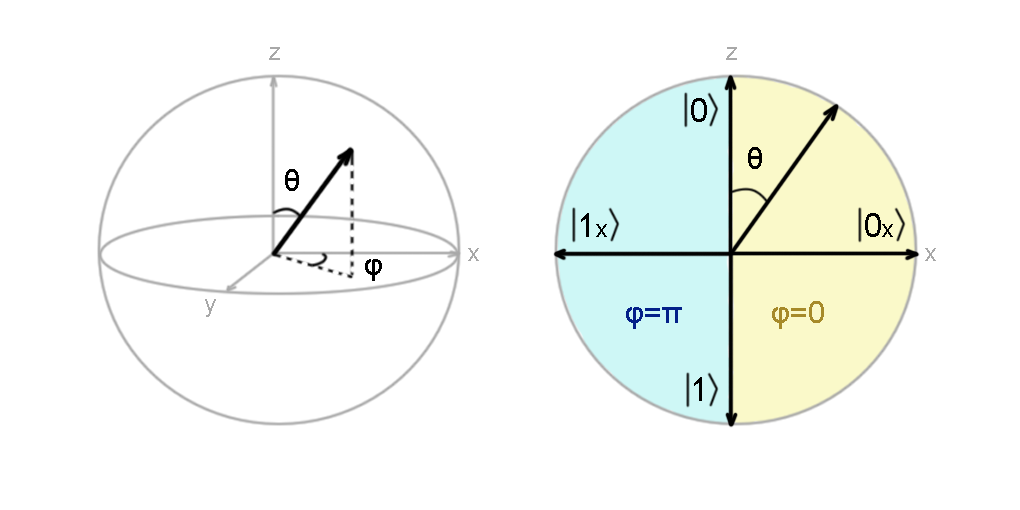}
\caption{(Left) Bloch representation of quantum states in terms of the polar $\theta$ and azimuthal $\varphi$ angle. (Right) $x-z$ plane of the Bloch sphere with $\varphi=0$ and $\varphi=\pi$. States $\ket{0}$, $\ket{1}$, $\ket{0_x}$ and $\ket{1_x}$ are depicted whose spherical angles $(\theta,\varphi)$ are $(0,0)$, $(\pi,0)$, $(\pi/2,0)$ and $(\pi/2,\pi)$, respectively.}\label{bloch sphere}
\end{figure}

We consider qubits, i.e., two-level quantum mechanical systems with one degree of freedom that can have two possible values which we denote $0$ and $1$. Physically, this degree of freedom can be, for instance, the spin of a particle or the polarization of a photon. One of the main features that distinguishes quantum states from classical ones is the superposition principle where the qubits can be in a superposition of the two states $0$ and $1$, i.e. $\ket{\psi}=\alpha\ket{0}+\beta\ket{1}$, where $\alpha$ and $\beta$ are complex numbers with $|\alpha|^2+|\beta|^2=1$. Thus, there exist infinitely many superposition states. All these superposition states can be easily visualized as unit vectors in the Bloch sphere \cite{Dur13,Dur14,Dur16}. In particular, the states $\ket{0}$ and $\ket{1}$ are the Bloch vectors pointing upwards and downwards, respectively, in the $z$ direction. The superposition states $\ket{0_x}=\frac{1}{\sqrt{2}}(\ket{0}+\ket{1})$, $\ket{1_x}=\frac{1}{\sqrt{2}}(\ket{0}-\ket{1})$ can be visualized as unit vectors pointing along the $x$ direction. In general, any unit vector of the Bloch sphere corresponds to a quantum state that can be determined by the two angles of the spherical coordinates (see Figure~\ref{bloch sphere} (Left)) as
\begin{equation}
\ket{\psi}=\cos\left(\frac{\theta}{2}\right)\ket{0}+e^{i\varphi}\sin\left(\frac{\theta}{2}\right)\ket{1}\label{psi bloch}
\end{equation}
where $\theta \in [0,\pi]$ and $\varphi \in [0,2\pi)$ are the polar and azimuthal angles, respectively.

For the purpose of this proposal it is enough to work only with one Bloch plane. In particular, we consider the $x-z$ plane with azimuthal angles $\varphi=0$ and $\varphi=\pi$ (see Figure~\ref{bloch sphere} (Right)). Given a generic direction $a$, separated from the $z$-axis by an angle $\theta_a$, the state $\ket{0_a}$ is determined by $\varphi=0$ (right side of the plane) and $\theta_a$, i.e. \mbox{$\ket{0_a}=\cos\left(\frac{\theta_a}{2}\right)\ket{0}+\sin\left(\frac{\theta_a}{2}\right)\ket{1}$}. Analogously, the state $\ket{1_a}$ is determined by $\varphi=\pi$ (left side of the plane) and $\theta_a$, i.e. \mbox{$\ket{1_a}=\cos\left(\frac{\theta_a}{2}\right)\ket{0}-\sin\left(\frac{\theta_a}{2}\right)\ket{1}$}. If this mathematical formalism is introduced at high school levels, one can alternatively consider only one real parameter $\theta \in [0,2\pi)$ in order to avoid complex numbers.

The next feature that characterizes qubits is the probabilistic nature of the measurement process. According to Born's rule, if one measures the state $\ket{\psi}$ given by $\ket{\psi}=\alpha\ket{0}+\beta\ket{1}$ in the $z$ direction, the probability of obtaining the state $\ket{0}$ ($\ket{1}$) is \mbox{$p_0=|\alpha|^2$} ($p_1=|\beta|^2$).  Therefore, one obtains a deterministic outcome after a measurement in $z$ direction, i.e. $\alpha=1$ ($\beta=1$), only if the initial state is $\ket{\psi}=\ket{0}$ ($\ket{\psi}=\ket{1}$). State $\ket{0}$ gives outcome $+1$, whereas state $\ket{1}$ gives outcome $-1$. In algebraic terms, these two states are called eigenstates of the $z$ basis with eigenvalues $+1$ and $-1$. Considering the representation of $\ket{\psi}$ in the Bloch sphere (Eq.~\eqref{psi bloch}), the probabilities $p_0$ and $p_1$ can be written in terms of the spherical angles $\theta$ and $\varphi$ as
\begin{eqnarray}
p_0&=&\cos^2 \frac{\theta}{2}, \label{prob theta1}\\
p_1&=&\sin^2 \frac{\theta}{2}. \label{prob theta2}
\end{eqnarray}
The probability of getting either the outcome $+1$ or the outcome $-1$ after a measurement in the $z$ direction only depends on the angle $\theta$, which is in this case the separation between the measurement direction ($z$-axis) and the Bloch vector of the state $\ket{\psi}$. The smaller the angle $\theta$, the higher the probability $p_0$ to obtain outcome $+1$. More generally, the measurement process can be performed in any direction $a$. Mathematically, the measurement is described by the observable $O_a$, of the form
\begin{equation}
O_a=(+1)\ket{0_a}\bra{0_a}+(-1)\ket{1_a}\bra{1_a},
\end{equation}
where $\ket{0_a}$ and $\ket{1_a}$ are the eigenstates of the $a$ basis, with eigenvalues (outcomes) $+1$ and $-1$, respectively, i.e. $O_a \ket{0_a}=(+1)\ket{0_a}$ and $O_a \ket{1_a}=(-1)\ket{1_a}$. Analogously to Eq.~\eqref{psi bloch}, one can express an initial state $\ket{\psi}$ in terms of the eigenstates $\{\ket{0_a}, \ket{1_a}\}$ of any basis $a$. Thus, the probability $p_0$ of getting the outcome $+1$ is given by the angle $\theta_{\psi,a}$ between the Bloch vector of the initial state $\ket{\psi}$ and the state $\ket{0_a}$, that specifies the measurement direction $a$, i.e. $p_0=\cos^2 \frac{\theta_{\psi,a}}{2}$.

Another crucial property of quantum mechanics is that, once the measurement process is performed, the initial state changes into one of the two eigenstates of the measurement direction $a$ depending on the outcome. For instance, if the outcome was $+1$, the final state is $\ket{0_a}$, and if it was $-1$, the final state is $\ket{1_a}$. Therefore, no more information can be accessed from the original state since the state is no longer $\ket{\psi}$.

\subsubsection{Rules of the game}\label{rules one qubit}
\begin{figure}[ht!]
\includegraphics[width=3.4in]{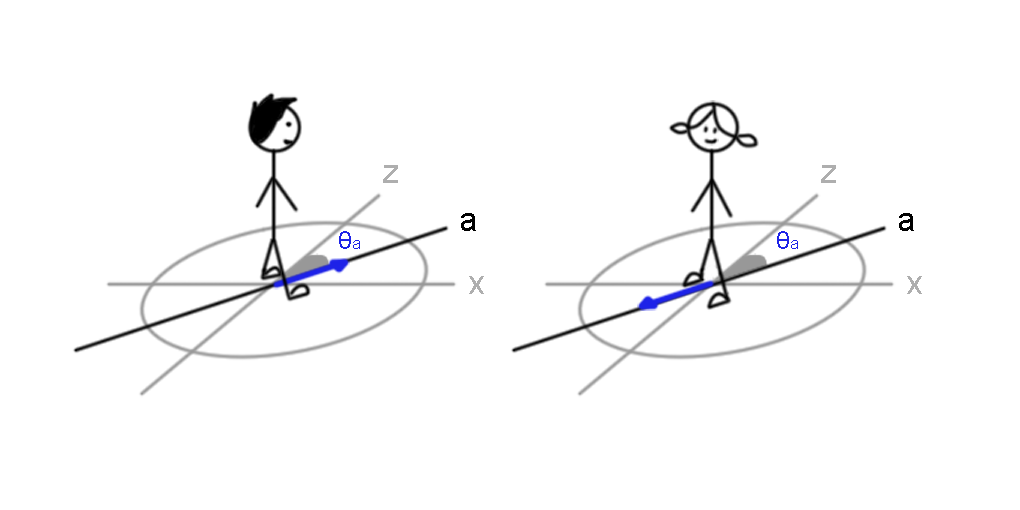}
\caption{The qubit-students themselves represent the Bloch vector in the $x-z$ Bloch plane painted on the floor. (Left) A student represents a qubit in the state $\ket{0_a}$ when they rotate by an angle $\theta_a$ from the vertical axis. (Right) The qubit-student faces towards the opposite side to represent the state $\ket{1_a}$ along the $a$-axis.}\label{student bloch vector}
\end{figure}

As described in Section~\ref{theo one qubit}, the first feature our game should reproduce is the qubit states. We only consider the states that can be described within the $x-z$ plane of the Bloch sphere and which allows us to represent it by drawing a circle on the floor. Once the Bloch plane is painted, the qubit-students mimic the Bloch vector with their body orientation as shown in Figure~\ref{student bloch vector}. As argued in Section~\ref{theo one qubit}, the angle $\theta$ is enough to describe states in this plane, where the angle $\varphi=0,\pi$ is only responsible for a phase. Therefore, given a generic direction $a$ with angle $\theta_a$ from the vertical axis, the qubit-student just needs to rotate their body by an angle $\theta_a$ to represent the state $\ket{0_a}$ (Figure~\ref{student bloch vector} (Left)). Already in this position, if the student turns around and faces the other side of the $a$-axis, they now represent the state $\ket{1_a}$ (Figure~\ref{student bloch vector} (Right)).

In addition to the states, we also need to reproduce the measurement process, which is stochastic due to the intrinsic nature of quantum mechanics. In order for the students to act stochastically the same way as qubits do, we propose that they use a biased die, which should be constructed in such a way that the probabilities $p_0$ and $p_1$ (Equations~\eqref{prob theta1} and \eqref{prob theta2}) can be obtained depending on the measurement directions used in the activity. For example, if the scientists are only allowed to prepare states and measure in the $x$ and $z$ directions, i.e. $p_0=p_1=1/2$ ($\theta=\pi/2$), a standard die suffices: odd numbers may indicate outcome $+1$ and even numbers outcome $-1$. In Sections \ref{bell test} and \ref{E91}, several activities are proposed that need other measurement directions and thus more examples of biased dice are given (see Appendix~\ref{app biased dice}).

\begin{figure}[ht!]
\includegraphics[width=3.4in]{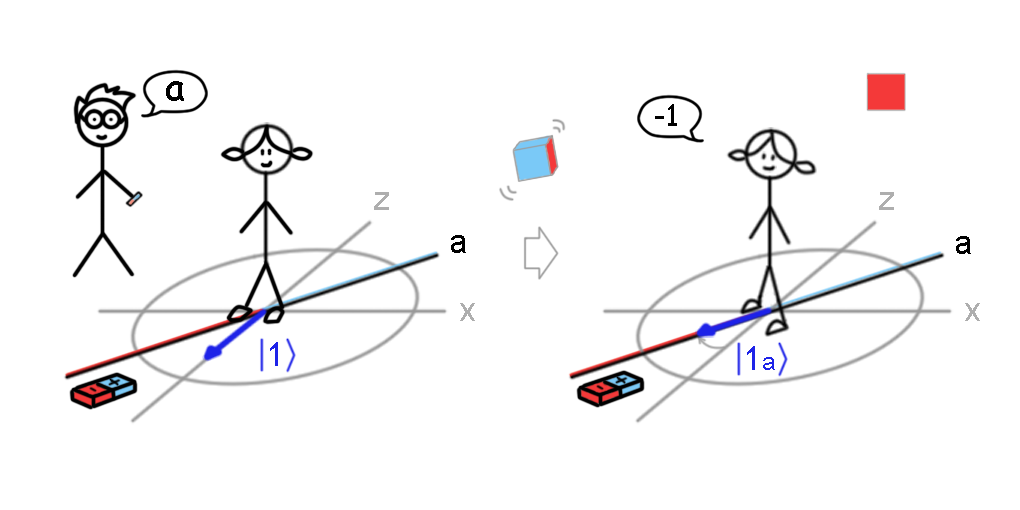}
\caption{Measurement process. After the qubit-student is hit with a ball, the scientist announces a measurement direction, places the compass stick to establish the "positive" side of the axis more clearly (see text for details), and the qubit acts accordingly. Here, the scientist announces a direction $a$ that is different from the one that the qubit-student is aligned with ($z$ in this case). Thus, the qubit-student throws a biased die, orients their body according to the die result and announces the outcome corresponding to their final state after the measurement. In the example shown, the die result is "Go to negative side" so the qubit-student aligns with the compass stick direction and faces towards the negative sign of the stick. Therefore, they announce the outcome $-1$.}\label{meas one par}
\end{figure}

The game proceeds as follows: the scientist prepares the qubits in an initial state. There are two possible states $\left\lbrace \ket{0_a}, \ket{1_a}\right\rbrace$ per measurement direction $a$. The qubit-students should orient their bodies with the angle that corresponds to the state the scientists chose. Then, the scientist starts to measure the qubits in different directions. A qubit is measured when the scientist hits them with a ball, thus simulating the real measurement process of e.g. a photon "hitting" a particle. If the scientist succeeds, they choose a measurement direction and place a \textit{compass stick} on the floor to make the direction more explicit and to mark which side of the axis is the "positive" side -- since it corresponds to the outcome $+1$ -- (see Figure~\ref{meas one par}). Once a qubit-student is hit, they have two possibilities:
\begin{enumerate}
\item If the qubit-student is already aligned with the measurement direction, they announce the outcome corresponding to their state, i.e., either $+1$ (if they are facing towards the positive sign of the compass stick) or $-1$ (if they are facing towards the negative sign of the compass sign).
\item If the qubit-student is pointing towards a different direction, then they throw the corresponding biased die and, depending on the result, orient towards the $\ket{0_a}$ (positive side of the $a$-axis) or $\ket{1_a}$ (negative side) state and announce the outcome (see Figure~\ref{meas one par} for an example).
\end{enumerate}

\subsection{Entangled pair of qubits}\label{entangled qubits}
\subsubsection{Theoretical background}
We now consider two-qubit states, in particular, entangled states. The main feature of entangled states is that the measurement outcomes of the two qubits are correlated in more than one measurement direction. One representative example of these entangled states is the so called Bell state $\ket{\psi^{-}}$ that reads
\begin{equation}
\ket{\psi^{-}}=\frac{1}{\sqrt{2}}(\ket{0}_A \otimes \ket{1}_B- \ket{1}_A \otimes \ket{0}_B)
\end{equation}
where subindex $A$ denotes the first qubit and subindex $B$ the second one. In our game, we deal with this particular state, where one scientist (let us call her Alice) measures qubit $A$ and another scientist (let us call him Bob) measures qubit $B$. In actual experiments, these two qubits can be arbitrarily separated in space and the measurement outcomes are still correlated, contrary to any (classical) local theory. As in the single-qubit case, one can apply Born's rule and notice that after Alice measures her qubit in the $z$ direction, she will get either the outcome $+1$ (state $\ket{0}_A\otimes\ket{1}_B$ after the measurement) with probability $1/2$ or the outcome $-1$ (state $\ket{1}_A\otimes\ket{0}_B$) with probability $1/2$. Therefore, when Bob measures his qubit in the same direction as Alice, the state that he gets is correlated to Alice's, i.e., if Alice gets the outcome $+1$, Bob gets outcome $-1$ and if Alice gets the outcome $-1$, Bob gets $+1$. This occurs not only in the $z$ direction as explained here, but also in any other direction\footnote{In an arbitrary direction $a$, this Bell state is written, up to a global phase, as \mbox{$\ket{\psi^-}=\frac{1}{\sqrt{2}}(\ket{0_{a}}_A\otimes\ket{1_{a}}_B-\ket{1_{a}}_A\otimes\ket{0_{a}}_B)$}.}, which is a genuinely quantum feature. Hence, measurement outcomes are always anti-correlated when both Alice and Bob measure the state $\ket{\psi^-}$ in the same direction.

\subsubsection{Rules of the game}\label{rules entangled}
In order to represent the entangled state $\ket{\psi^-}$, the students playing the role of qubits hold hands facing each other and start spinning at the center of the Bloch plane on the floor, which symbolizes the fact that they are entangled in all directions.

\begin{figure}
\includegraphics[width=3.4in]{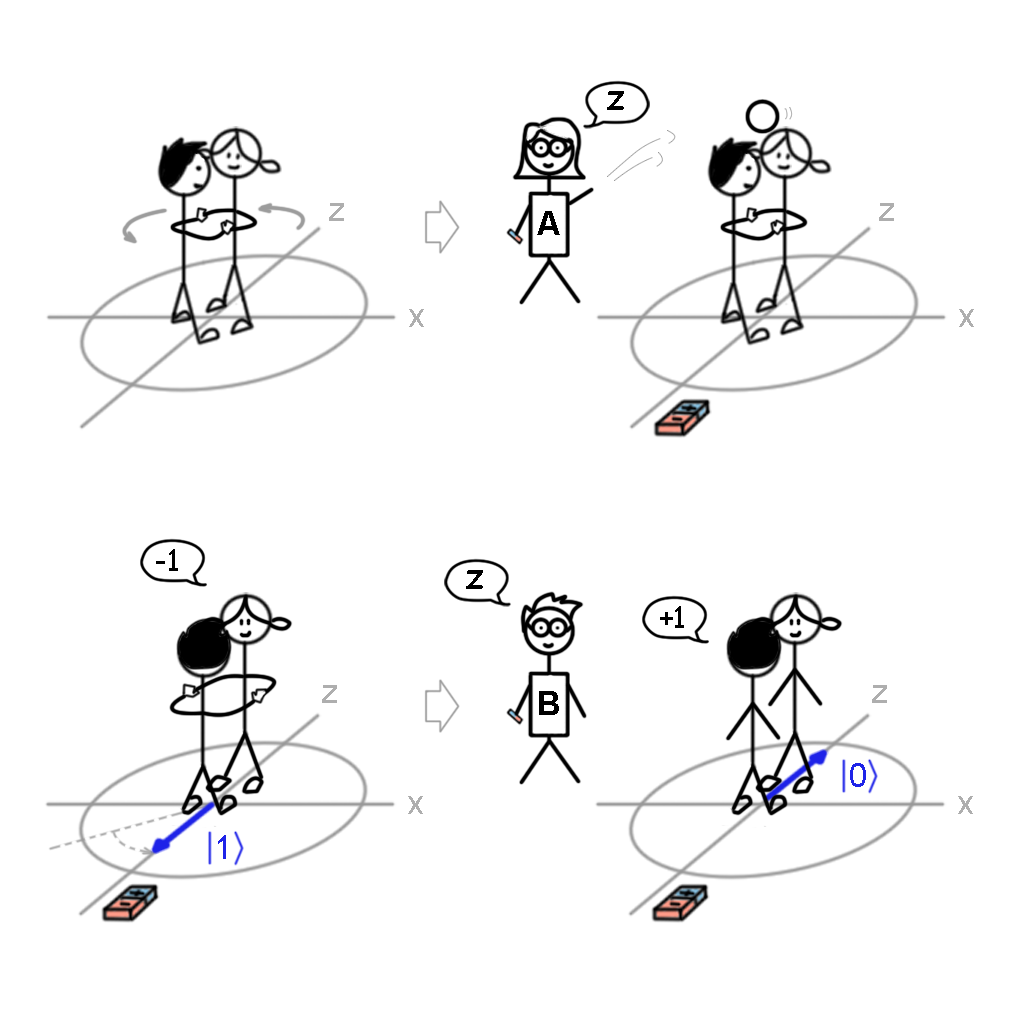}
\caption{(Up Left) Representation of an entangled state $\ket{\psi^-}$ by two students that hold hands and spin around the center of the Bloch plane. (Up right) Alice measures one of the qubits of the entangled pair. She announces the measurement direction and marks it with the compass stick on the floor, so that the positive side and the negative side of the axis are clear. Then, she tries to hit her qubit with the ball. (Down left) When Alice succeeds, both qubits keep rotating until they reach the measurement direction marked with the compass stick (in this case, the $z$-axis), and the measured qubit announces the corresponding result $-1$ (negative side). (Down Right) Bob measures the other qubit of the pair in the same direction as Alice and obtains the anti-correlated result $+1$, since the qubit is facing opposite to Alice's qubit. Note that qubits are no longer entangled after Alice's measurement.}\label{entangled pair}
\end{figure}

The measurement process is illustrated in Figure~\ref{entangled pair}. When the scientist that plays the role of Alice measures one of the qubits, she announces the measurement direction, places the compass stick and tries to hit the qubit with the ball.\footnote{Note that in this case, the scientist chooses a measurement direction before they actually managed to hit the qubit-student. Even if this is not completely realistic, we consider that this order is clearer for the qubit-students, who have to rotate until they get to the compass stick when measured. Placing the compass stick after the hitting-with-ball process would just increase the difficulty for the qubit students to perform in the correct way.} Since the two qubits are holding hands, once one qubit is hit, the whole pair keeps rotating until they get to the direction marked by the compass stick. Alice's qubit then announces out loud the outcome, which is $+1$ if she is facing towards the positive sign of the compass stick and $-1$ otherwise. Note that each of these outcomes has probability 1/2 to occur for Alice's qubit, since the entangled pair is spinning when one of the qubit-students is hit, which can happen randomly at any point of the rotation, leading to an equal probability to get first to the positive or to the negative sign of the stick.  If a second scientist (Bob) measures now the other qubit of the entangled pair, the latter behaves as a normal single qubit, i.e. as explained in Section~\ref{rules one qubit}, since the measurement of Alice's qubit has resulted in both qubits changing their initial state. In particular, if Bob measures in the same direction as Alice, it is clear that he obtains the anti-correlated result\footnote{If Alice gets $+1$, Bob gets $-1$ and vice versa.} since Bob's qubit is already facing opposite to Alice's (see Figure~\ref{entangled pair}). Note that after the measurement, the qubits are no longer entangled (they do not hold hands anymore), so any further measurement does not lead to anti-correlated results, but to the qubits behaving independently of each other.

These basic rules introduced in Sections~\ref{one qubit} and~\ref{entangled qubits} allow for several games to be played. One possibility is to play the game as introduced in \citep{LopezDur2019}, where only the qubits know the rules they have to follow and the scientists should figure them out by making measurements and analyzing the results. This game thus focuses on the fundamental concepts of quantum mechanics (superposition states, measurement processes, Heisenberg's uncertainty relation, entanglement, etc.).

The other possibility is to play the cryptography games (see Sections~\ref{BB84} and~\ref{E91}), which consist in performing the BB84 and the E91 protocols for quantum key distribution. In both cases, there are two scientists (or teams of scientists) whose goal is to share a secret message. They should use the qubits to create and share a secure key to encrypt and decrypt the message, so that nobody else apart from them has access to it.

\section{Bell Test}\label{bell test}
The proposal presented in Section~\ref{entangled qubits} allows one to measure the qubits in any direction and to work with entangled qubits, in particular with those qubits in the Bell state $\ket{\psi^-}$. Thus, we already have all the necessary tools to mimic a Bell test.

The Bell test is an experiment designed to test Bell's theorem, introduced by John S. Bell \cite{RevModPhys.86.419, PhysicsPhysiqueFizika.1.195, RevModPhys.38.447}. Bell's theorem was a response to the Einstein-Podolsky-Rosen (EPR) paradox \cite{PhysRev.47.777} which was ought to show that the mathematical description of quantum mechanics is incomplete. EPR argued that the probabilistic nature comes from our ignorance of some degrees of freedom, called hidden variables, that we cannot access. The argument in the EPR paradox is based on the assumption of local realism, i.e., the state of one system cannot influence the state of a separated system instantaneously (locality) and the systems have fixed properties, no matter if the properties are measured or not (realism). On the other hand, Bell's theorem states that no local, realistic theory can reproduce all the predictions of quantum mechanics. More precisely, Bell's theorem puts a bound on the strength of correlations between the results of measurements performed on spatially separated systems, under the assumption that local realism holds. In turn, quantum mechanics predicts stronger correlations, which hence implies that one (or both of the assumptions of Bell's theorem - namely locality or realism - are wrong. Recent experiments have indeed shown that nature behaves in the way quantum mechanics predicts, i.e., there have been loophole free tests of Bell's inequality \cite{Bell1, PhysRevLett.115.250402, PhysRevLett.115.250401} that show a clear violation. It should be stressed though that Bell inequalities are not only about quantum mechanics, but they are far more general. They apply to {\it all} local realist theories, and if we find in experiments that Nature indeed violates these inqualities, this shows on the one hand that the predictions of quantum mechanics seem to be correct, and that quantum mechanics is indeed a theory where either reality or locality are not respected. What is more, it means that any future theory that may eventually replace quantum mechanics cannot be local realistic either - Nature simply does not behave this way! This is what makes Bell's theorem so interesting also from a fundamental point of view, as it tells us something about the fundamental functionality of our world.
     
In the following, we present two different approaches of testing Bell's theorem in class; these two approaches can be performed within the framework of our proposal.

\subsection{CHSH Inequality}\label{sectionCHSH}
\subsubsection{Theoretical background}
One approach to experimentally test Bell's theorem is to check the CHSH inequality \cite{PhysRevLett.23.880}
\begin{equation}
|\langle ab\rangle - \langle ab'\rangle + \langle a'b\rangle  + \langle a'b'\rangle | \leq 2, \label{eqCHSH}
\end{equation}
where $\langle \bullet \rangle$ denotes the expectation value operation and $a, b, a', b'$ are observables that have two possible measurement outcomes $\{\pm 1\}$ (see Appendix~\ref{app CHSH} for more details).

Let us explain the details of this inequality in order to understand the implications of its violation. Let us consider two systems, one in Alice's laboratory and the other one in Bob's. In this situation, $a$ and $a'$ denote two possible measurements that Alice can choose. Analogously, Bob can measure either $b$ or $b'$. If one assumes that the systems have fixed properties before any measurement (realism), there are two possible situations for Alice's outcomes (considering that the two possible outcomes are $\pm 1$):
\begin{eqnarray}
a'&\neq& a:\,\,\, a'+a=0,\,\,\,\,\,\,\, a'-a=\pm 2, \\
a'&=&a:\,\,\, a'+a=\pm 2,\,\,\, a'-a=0.
\end{eqnarray}
If one also considers the results of Bob's measurements, the quantity for S reads
\begin{equation}
S=(a'+a)b+(a'-a)b'=\pm 2,
\end{equation}
for both situations $a=a'$ and $a \neq a'$ with $b, b' \in \{\pm 1\}$. Using the triangle inequality, it is easy to see that
\begin{equation}
|\langle S \rangle | \leq \langle |S| \rangle=2, \label{eq:S}
\end{equation}
which is the CHSH inequality in Eq.~\eqref{eqCHSH}. Thus, if a theory violates this inequality, the initial assumptions of either locality or realism or both should be wrong. Therefore, only by making measurements in different directions and averaging the results, one can experimentally test a theoretical hypothesis such as local realism.

It has been shown (see Appendix \ref{app CHSH} for details) that quantum mechanics violates the CHSH inequality. In particular, the maximal violation ($|\langle S_\mathrm{QM} \rangle |=2\sqrt{2}$) occurs in the case where Alice and Bob share a pair of maximally entangled qubits (for example in the state $\ket{\psi^-}$) and the measurement directions $a$, $b$, $a'$ and $b'$ are in the same plane and successively separated by an angle $\pi/4$ (see figure~\ref{fig direc CHSH}). This experiment is precisely what one can perform in class with this proposal. Students can measure pairs of entangled qubits in any measurement direction, as described in Section~\ref{entangled qubits} (see also Figure~\ref{entangled pair}). Alice and Bob perform measurements in two different directions each of them can choose -- denoted ($a,a'$) for Alice and ($b,b'$) for Bob -- and write down the results. In order to check if the CHSH inequality is violated, they take the outcomes of the cases where the two measurement directions do not coincide and compute the expectation values in Eq.~\eqref{eqCHSH} afterwards.

\subsubsection{Rules of the game}\label{rules CHSH}
\begin{figure}
\subfigure[Measurement directions\label{fig direc CHSH}]{\includegraphics[width=3.4in]{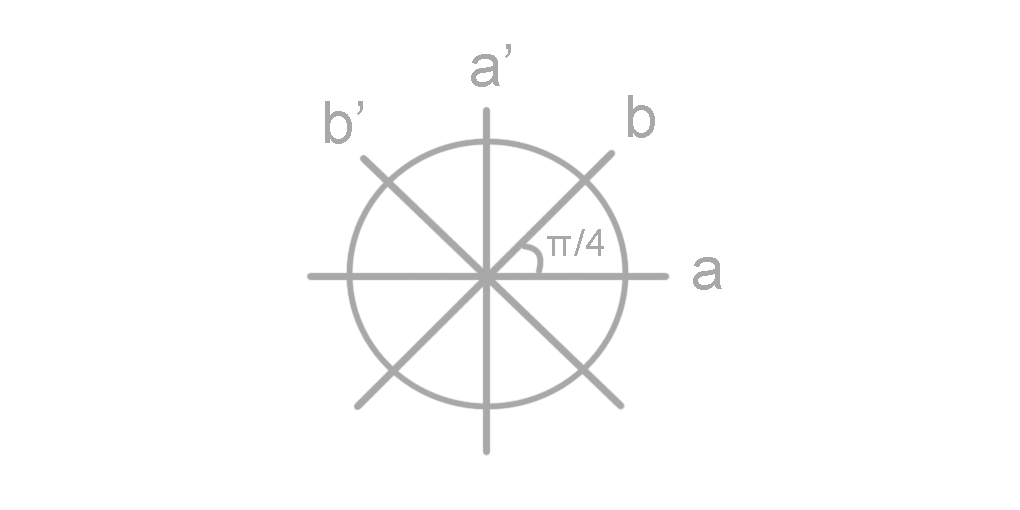}}
\subfigure[Performance of the game\label{Alice meas CHSH}]{\includegraphics[width=3.4in]{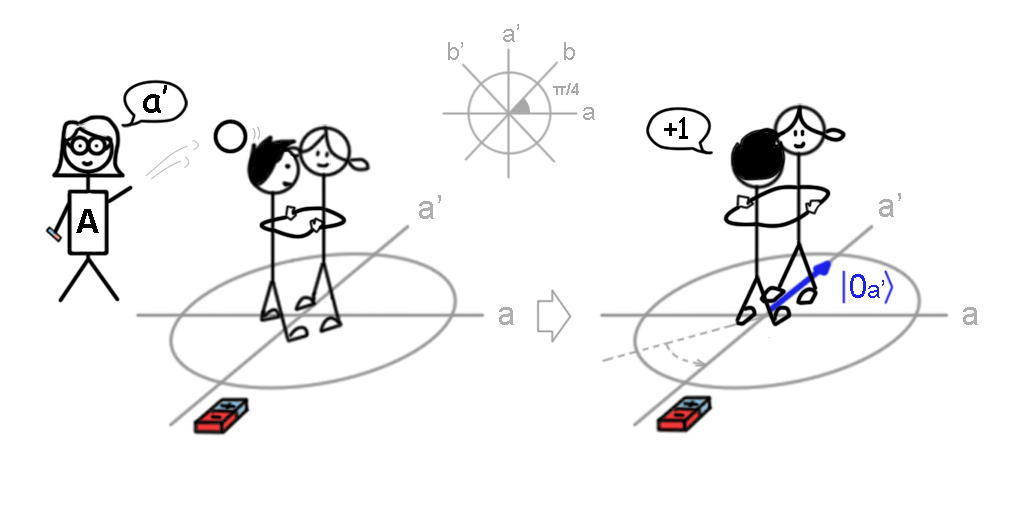}}
\caption{[a] Measurement directions $a$, $b$, $a'$ and $b'$ are in the same plane and successively separated by an angle $\pi/4$. [b] (Left) Alice measures her qubit in the $a'$-direction. The compass stick on the floor indicates the orientation of $a'$-axis. Once the qubit is hit with the ball, the pair rotates until they reach the $a'$-axis. (Right) Alice's qubit communicates the outcome out loud ($+1$ in the picture --aligned with the compass stick--, corresponding to the state $\ket{0_{a'}}$ in this case. Note that the initial state $\ket{\psi^-}$ becomes $\ket{0_{a'}}_A\otimes\ket{1_{a'}}_B$ after the measurement).}
\end{figure}

\begin{figure}
\includegraphics[width=3.4in]{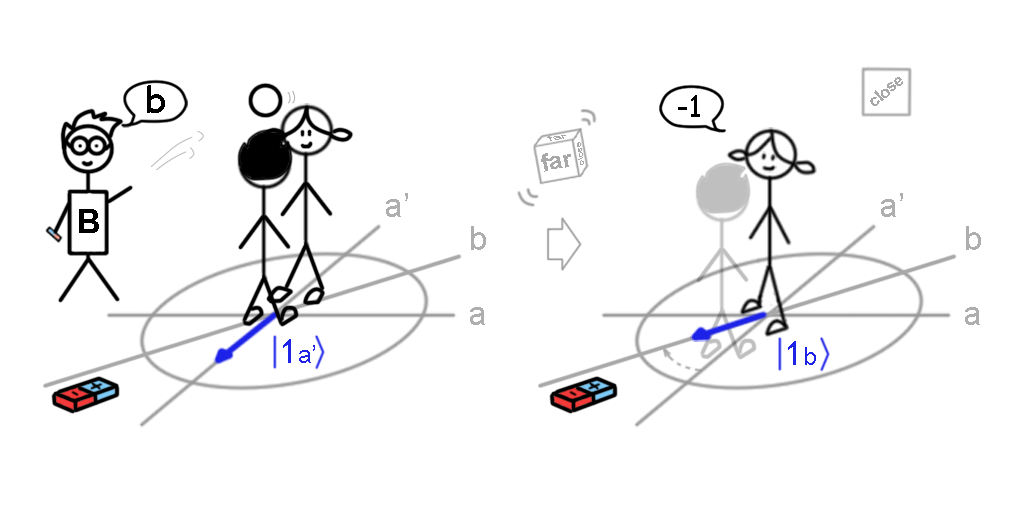}
\caption{(Left) Bob measures his qubit after Alice's measurement. The state of Bob's qubit after Alice's measurement is $\ket{1_{a'}}$. Since Bob measures in the $b$-direction, which is separated $\pi/4$ from the $a'$-axis, Bob's qubit has to throw a biased die to determine how it rotates. In this case, the die is such that, with probability $p_1 \approx 0.85$, the qubit rotates to the negative side of the $b$-axis (indicated with the negative sign of the compass stick) and with probability $p_0 \approx 0.15$, the qubit rotates to the positive side (see main text for details). (Right) After throwing the die, Bob's qubit obtains "Go to the closer side", so it rotates to the negative side in this case and communicates the outcome out loud to Bob. In this example, the final state for Bob's qubit is $\ket{1_b}$.}\label{Bob meas CHSH}
\end{figure}

In this section, we explain how Alice and Bob measure the state $\ket{\psi^-}$ in directions separated by an angle $\pi/4$ (see figure~\ref{fig direc CHSH}) using our proposal of Section~\ref{entangled qubits}. Let us take the example of directions $a'$ and $b$ whose measurement results are used to compute $\langle a'b\rangle$ in Eq.~\eqref{eqCHSH}. First, Alice chooses the measurement direction $a'$, places the compass stick in the orientation she wants, and hits her qubit-student with the ball (see Figure~\ref{Alice meas CHSH} (Left)). The entangled pair which was spinning before being hit by the ball, rotates until they reach the $a'$-axis and stops there. Then, Alice's qubit communicates the outcome out loud, i.e. $+1$ if he is facing towards the positive sign of the compass stick and $-1$ otherwise (see Figure~\ref{Alice meas CHSH} (Right)). The initial state $\ket{\psi^-}$ becomes either $\ket{0_{a'}}_A\otimes\ket{1_{a'}}_B$ (outcome $+1$) or $\ket{1_{a'}}_A\otimes\ket{0_{a'}}_B$ (outcome $-1$) after the measurement. In the example of Figure~\ref{Alice meas CHSH}, the final state is $\ket{0_{a'}}_A\otimes\ket{1_{a'}}_B$. Now, Bob measures his qubit in direction $b$, so he places his stick to mark it (see Figure~\ref{Bob meas CHSH} (Left)). Note that his qubit has changed its state due to Alice's measurement. In the example of Figure~\ref{Bob meas CHSH}, the qubit is in the state $\ket{1_{a'}}_B$. Since the measurement direction is rotated by $\pi/4$ from the initial state direction $\ket{1_{a'}}_B$, Bob's qubit has to throw the biased die to decide if she rotates to the positive or to the negative side of axis $b$ (Figure~\ref{Bob meas CHSH}). The biased die should be designed such that, with probability $p_0=\cos^2 \left(3\pi/8\right)\approx 0.15$, the qubit-student rotates to the positive side of the $b$-axis which is indicated by the positive sign of the compass stick; and with probability $p_1=\sin^2 \left(3\pi/8\right)\approx 0.85$, the qubit-student rotates to the negative side of the $b$-axis (see Figure~\ref{Bob meas CHSH} (Left)). From this example, it is intuitively easy to see that the student has a higher probability of rotating to the side of the measurement axis that is closer to them. With this intuitive idea, the die's possible events can be "Go to the closer side" or "Go to the further side", which are completely general events that cover all possibilities.  In Appendix~\ref{app biased dice}, several examples of biased dice are given for different separation angles between the measurement directions $a, b, a'$ and $b'$. To mimic the corresponding probabilities in each case, we propose two alternatives to construct the dice, either to physically build $k$-sided dice or to make use of apps that allow one to choose the number of sides. The creation of the dice can also be considered as an activity itself to work with probabilities.

Once Bob's qubit has rotated according to the die result, the corresponding qubit-student communicates the outcome out loud to Bob (Figure~\ref{Bob meas CHSH} (Right)). Alice and Bob should repeat this measurement process with as many entangled pairs as possible in order to get a statistical reliable value of $\langle a'b\rangle$. Then, the same procedure should be done also for the combinations $\langle ab\rangle$, $\langle ab'\rangle$ and $\langle a'b'\rangle$. In Appendix~\ref{app stat error}, we present some estimations of the number of measurements that Alice and Bob should do as well as estimations of the number of students needed to perform the whole process in order to get a statistical accurate value for the quantity $|\langle S \rangle |$ of Eq.~\eqref{eq:S}.

In order to make a clear description of the game, we have explained the measurement process in such a way that Alice measures first and Bob measures second, but this order can be changed without loss of generality or changing the results. We also remark that a crucial property of real entangled particles is that they give correlated outcomes no matter how far away they are from each other, so Alice's and Bob's laboratories do not need to be in the same place. The game as we present it here does not mimic this property since the entangled students hold hands, which forces them to be in the same place. A possible variant of the game could be introduced that includes such a property, namely the entangled students could imitate each other and perform as explained above while they are placed in Bloch circles that are separated from each other. In this case, if one of the qubit rotates, the other one also rotates and when one is measured, both keep rotating until they reach the measurement axis marked on the Bloch plane of the measured student\footnote{In this situation (entangled qubits are placed in two different Bloch circles), if both qubits are hit at the same time by the ball, they stop spinning and rotate until both of them get to the measurement direction chosen by the scientist who measures first.}.

Although the activity can be presented to the students in numerous ways, we suggest to use it as described here to enhance their critical thinking. Thus, we propose the following unit (we assume that the initial concepts of superposition and entanglement have been already taught beforehand and we focus only on the Bell test):
\begin{enumerate}
\item The ideas of locality and realism are presented. The students should discuss their own initial concept of reality based on the concepts of locality and realism.
\item Considering the initial discussion, the dichotomous outcomes are presented and the students should work with the expression for the quantity $S$. Then, the CHSH inequality is presented.
\item The students should perform all the necessary measurements on the entangled pairs in order to check the CHSH inequality (this part is performed with the game introduced above).
\item After gathering all the experimental data, students should compare their initial intuition with the resulting data and analyze the implications of the violation of the CHSH inequality.
\end{enumerate}

In addition to the CHSH approach, we also propose an alternative to test local realism. It is based on the work "Is the moon there when nobody looks? Reality and the Quantum theory", by N. David Mermin \cite{Mermin}. This approach relies less than the CHSH inequality on the mathematical background of the students, so that one can directly apply it without any previous explanation. However, more measurements are needed to get good statistics, which translates into larger groups of students needed (see Appendix~\ref{app stat error} for a detailed estimation on the number of students and measurements needed). The details of this alternative and a comparison to the CHSH approach are presented in Appendix~\ref{Mermin}.

\section{Application: Quantum Cryptography}\label{QuantumCryptography}
Considering our proposals from the previous sections, one may also use them to make a feasible approach to quantum cryptography, which is a direct application of key concepts in quantum mechanics such as superposition states or entanglement. In this section, we show that the main protocols in quantum cryptography can be simulated in a realistic way with the tools presented in Section~\ref{Main proposal}.

\subsection{BB84}\label{BB84}
One of the most important cryptography protocols, which rely on the properties of quantum mechanics instead of the complexity of a mathematical problem, was proposed originally by G. H. Bennett and G. Brassard in 1984 and celebrated 30 years afterwards  \cite{Bennett_2014}.

\subsubsection{Theoretical background}\label{theoBB84}
In the BB84 protocol, a sender (Alice) wants to encrypt a message and send it to a receiver (Bob) via a secure channel. This is done by generating a private key (password that consists of a sequence of bits to encrypt the message) that only Alice and Bob know. The protocol is designed to generate this key, whose security relies on the properties of quantum mechanics. It proceeds as follows: Alice and Bob can send and receive their qubits in two bases, e.g. $z$ and $x$, which gives the set $\{ \ket{0}, \ket{1}, \ket{0_x}, \ket{1_x} \}$ of states in these bases. Here, $\ket{0}$ and $\ket{1}$ are the eigenstates in the $z$ basis and $\ket{0_x}=\dfrac{1}{2} (\ket{0}+\ket{1})$ and $\ket{1_x}=\dfrac{1}{2} (\ket{0}-\ket{1})$ the eigenstates in the $x$ basis.
The sender Alice sends the eigenstate of her qubit in a chosen basis to the receiver Bob, who chooses then in which basis he measures this received qubit. After this, Alice and Bob announce their chosen basis via a public channel. Alice and Bob repeat this procedure multiple times and then collect the results of the qubits for which the sent and measured basis coincide (while the other results with different basis are neglected). This is the quantum key with which Alice and Bob can encrypt messages. An intuitive example of applying the BB84 protocol and generating a secure quantum key can be seen in Table \ref{TableBB84}.

\begin{table}[ht]
\begin{center}
\begin{tabular}{p{1.5cm}p{0.6cm}p{0.6cm}p{0.6cm}p{0.6cm}p{0.6cm}p{0.6cm}p{0.6cm}p{0.6cm}lllllllll} \hline
\rule{0pt}{3ex}
A: & 0 & 1 & 1 &0 & 1 & 0 & 0 & 1\\
\rule{0pt}{3ex}
A basis: & z & z & x & z & x & x & x & z \\
\rule{0pt}{3ex}
B basis: & z & x & x & x & z & x & z & z \\
\rule{0pt}{3ex}
B: & 0 & 0 & 1 & 0 & 0 & 0 & 1 & 1\\[0.1cm]
\hline
\hline
\rule{0pt}{3ex}
Key: & 0 & & 1 & & & 0 & & 1
\end{tabular}
\end{center}
\caption{Example of generating a secure key based on quantum cryptography via the BB84 protocol.}
\label{TableBB84}
\end{table}

\subsubsection{Generating a key}\label{send secret message}
As we have shown in the previous Section~\ref{theoBB84}, only the four states $\{ \ket{0}, \ket{1}, \ket{0_x}, \ket{1_x} \}$ and the two measurement directions $x$ and $z$ are needed to reproduce the BB84 protocol.

In this section, we propose a specific activity that relies on the rules of the game explained in Section~\ref{rules one qubit}. The goal of the activity is to communicate a secret message successfully. To do so, the students form groups, each of which consists of one student playing the role of Alice, another one playing Bob and the rest will be qubits. First, Alice and Bob generate the key for the encryption of the message using the BB84 protocol. Then, Alice chooses the message she wants to send, encrypts it with that key and sends it to Bob. Finally, Bob uses the key to decrypt the message. They win if the message is transmitted successfully. Let us illustrate this activity with an example.

In order to make it easier, let us assign a number to each letter of the alphabet, i.e. A-1, B-2, ..., Z-26. If these numbers are written in binary, each letter will correspond to a binary code of 5 digits, e.g. $E=(00101)$. In our example, Alice wants to send the message "EY" ($E=00101, Y=11001$ in binary) to Bob. First, she needs to encrypt it with a password or \textit{key}, so that only a person that knows the key will be able to decrypt it and get the message. One simple way to encrypt it is by summing the message and the key in binary (mod 2)\footnote{$1\oplus1=0$, $0\oplus0=0$, $0\oplus1=1$, $1\oplus0=1$} as shown for example in Table \ref{Tableex}. This method of encryption is called one-time pad and it is designed in such a way that the encrypted message cannot be directly (without knowing the key) deciphered by taking advantage of e.g. language patterns such as the high frequency of certain letters. In particular, if the key is only used once, it is at least as long as the message and it is truly random, the resulting encryption after the sum will also be random (thus no patterns can be extracted). Therefore, Alice and Bob just need to ensure that only the two of them know the key in order to be sure that the message remains secret. Once both have the key, Alice can publicly send the encrypted message and only Bob will be able to decrypt it (he just needs to sum [mod 2] the key and the encrypted message to get the real message). 

\begin{table}[ht]
\begin{center}
\begin{tabular}{p{2cm}p{0.3cm}p{0.3cm}p{0.3cm}p{0.3cm}p{0.3cm}p{0.3cm}p{0.3cm}p{0.3cm}p{0.3cm}p{0.3cm}lllllllllll} \hline
\rule{0pt}{3ex}
Message: & 0 & 0 & 1 & 0 & 1 & 1 & 1 & 0 & 0 & 1\\
\rule{0pt}{3ex}
Key: & 1 & 0 & 0 & 1 & 1 & 0 & 0 & 0 & 0 & 1\\
\hline
\hline
\rule{0pt}{3ex}
Encryption: & 1 & 0 & 1 & 1 & 0 & 1 & 1 & 0 & 0 & 0
\end{tabular}
\end{center}
\caption{Example of message encryption.}
\label{Tableex}
\end{table}

Let us illustrate how Alice and Bob generate the key with the BB84 protocol according to the rules of our game.

\begin{figure}[ht!]
\includegraphics[width=3.4in]{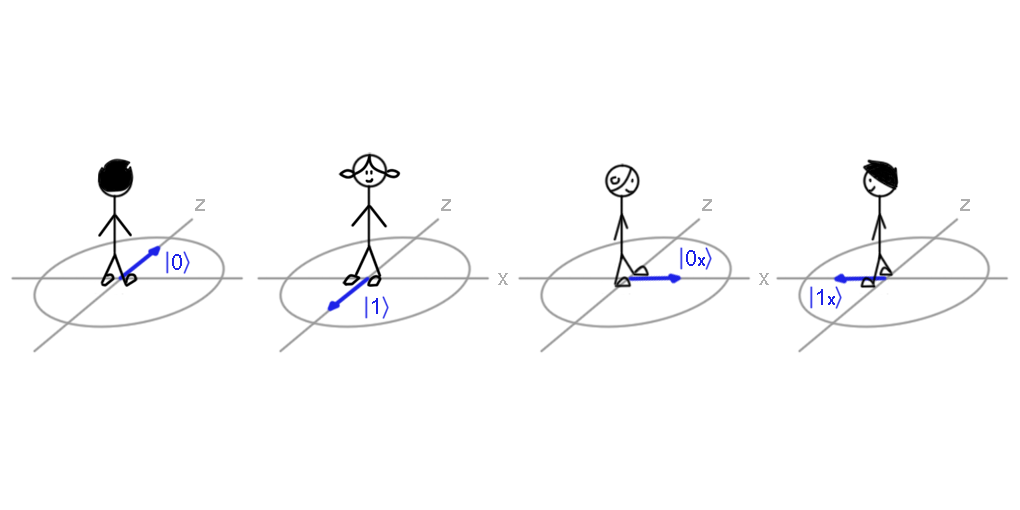}
\caption{Students' positions corresponding to the states $\{ \ket{0}, \ket{1}, \ket{0_x}, \ket{1_x} \}$ respectively (from left to right).}\label{students pos bb84}
\end{figure}

As explained in Section~\ref{theoBB84}, Alice has to prepare the qubits in certain states that Bob will afterwards measure. Since only two measurement directions are needed, the Bloch plane painted on the floor only needs to have two perpendicular axes (vertical for the $z$-axis and horizontal for the $x$-axis). Before the protocol starts, Alice and Bob should agree on which is the positive side of each axis and keep the same criterion throughout the whole protocol. In order for the qubits to know this criterion and for them to remember it, they can place compass sticks besides the axis on the floor (see Figure~\ref{meas one par}). Then, the protocol begins. Alice can prepare the qubits in the states $\{ \ket{0}, \ket{1}, \ket{0_x}, \ket{1_x} \}$, whose corresponding body orientations are depicted in Figure~\ref{students pos bb84}.

Bob measures Alice's qubits by following the rules explained in Section~\ref{rules one qubit}. In this case, a normal die is sufficient (odd numbers may indicate outcome +1 and even numbers outcome -1) to determine the qubit's orientation in the measurement process (see Table~\ref{Table:dices} in Appendix~\ref{app biased dice}). Bob writes down both the measurement direction he chose and the outcome he obtained, in order to create a table analogous to Table~\ref{TableBB84}. Note that Alice should also write down the prepared state (both the axis and the side of the axis towards which student is oriented). However, in order to create a key which is a string of binary digits (0 or 1), one cannot use directly the outcomes $+1$ and $-1$ because they are not binary. A simple transformation is enough to solve this: whenever the outcome is $+1$, the bit is $0$ (in analogy to $\ket{0}$ and $\ket{0_x}$, which are the states corresponding to the outcome $+1$); whereas the outcome $-1$ becomes the bit $1$.

Once Bob finishes measuring the qubits Alice prepares, both say the list of preparation and measurement axes (but not the bits!) out loud --i.e. they share it through a public channel--. Then, Bob checks which of those coincide with his measurement directions and take the corresponding outcome bits as the key for decryption.\footnote{Note that only half of the directions will coincide, so Alice should prepare at least twice as many qubits as the number of digits of the message.} Once the key is ready, Alice encrypts her message with it and sends it to Bob publicly (she says the encrypted digit string out loud). Bob decrypts the message with the key and they win the game if the message has been communicated successfully.

\subsubsection{Eavesdropping}
So far we have presented an illustrative example of just the key generation and the process of encryption and decryption. However, this is not a realistic scenario since the main purpose of a secure cryptography protocol is to be able to detect and prevent other parties from knowing the message. Thus, in this section we assume that there \textit{is} an Eavesdropper that tries to intercept the message Alice sends to Bob and we propose some strategies to detect it.

One of the main features of the BB84 protocol is that it provides an easy way, based on the quantum properties of qubits, to detect interventions in the key generation. As we have seen in Section~\ref{one qubit}, the initial state of quantum systems may change after the measurement process, which can be used to check if a third party has measured the qubits to extract information about the secret key.

The key --established following the BB84 protocol-- is generated using only the outcomes that Bob obtained when he measured in the same direction as Alice's preparation axis. This is due to the fact that the eigenstates of a measurement basis give a deterministic outcome (one obtains this outcome with probability 1) and they are the only states that do not change after the measurement process (see Section~\ref{theo one qubit}). Therefore, Alice and Bob can be sure that both share the same key bits whenever they worked with the same axis. This same property can also be used to detect the action of the Eavesdropper. Alice and Bob just need to use part of the key that they generate to check that both of them have the same bits in the same positions.\footnote{In order to compare the subset of bits, they can use a public channel (say it out loud in our game). Note that this subset can no longer be used to encrypt the message.} If this is so, the key is secure. If not, they abort the protocol and start the process again. However, since the measurement process is stochastic, the Eavesdropper has some probability to go unnoticed. For the sake of argument, let us assume for now that the Eavesdropper gets access to Alice's qubit. If, by chance, the Eavesdropper measures in the same direction as Alice's preparation axis, the state does not change at all, so there is no way to detect the external intervention. On the other hand, if the Eavesdropper's measurement direction does not coincide with Alice's, the prepared state is modified and Bob obtains a different outcome with 1/2 probability.\footnote{An example of this situation is the following: Alice prepares the state $\ket{0_x}$ in the $x$ direction. Then, the Eavesdropper chooses to measure in the $z$ direction, which transforms the qubit state into either $\ket{0}$ or $\ket{1}$. In both cases, if Bob then measures the qubit in Alice's direction ($x$), he does not obtain the state $\ket{0_x}$ deterministically --as it would be the case without the external intervention--, but with 1/2 probability. This is so because Bob's measurement direction ($x$) and the direction of the state he gets ($z$) are separated $\theta=\pi/2$, so the probability that he obtains the outcome $+1$ is $p_0=1/2$, given by Eq.~\eqref{prob theta1}.} In total, the Eavesdropper is detected with probability 1/4 at each measurement of one qubit (one bit of the key). Therefore, the more bits are used to check the security of the key, the less probable it is that the Eavesdropper goes unnoticed. The phase of checking described above will be referred to as \textit{security test} in the following.

In order to play the full version of the BB84 protocol, we propose that some of the students in each group play the role of Eavesdroppers and they have to come up with strategies to intercept the message without being discovered. Then, the goal of the game is for them to discover the key; and for Alice and Bob to send the message successfully and secretly.

We now propose a few possible eavesdropping strategies in case the teacher wants to suggest them to the students in advance. The preparation of states and the measurement processes are always performed following the rules of the game detailed in Section~\ref{rules one qubit}.

\textbf{1. The Eavesdroppers measure the qubit before Bob.} One way of getting the key is to intercept Alice's qubit, measure it and send it on to Bob so that he does not notice any difference. Equivalently in our game, the Eavesdroppers try to hit Alice's qubit with the ball before Bob but, since the qubit student does not move away from the Bloch circle, Bob can measure it after the Eavesdroppers. Note that the state of the qubit may have changed after the Eavesdroppers' measurement, depending on which basis they choose.

The Eavesdroppers proceed as Bob, i.e. they measure the qubit in one direction ($x$ or $z$) and then, when the list of preparation axes is made public by Alice and Bob, they only take the measurement outcomes of the coincident axes. In this way, the Eavesdroppers have access to the key and can decrypt the secret message. However, they get the correct bits with probability\footnote{Sum of the probabilities that Eavesdroppers choose the same direction as Alice and Bob and that they choose the other direction but still they get the same outcome, i.e. $\frac{1}{2}+\frac{1}{2}\cdot \frac{1}{2}=\frac{3}{4}$.} 3/4.

Alice and Bob detect the action of the Eavesdroppers with probability 1/4 per checked bit. Thus, Alice and Bob have to compare several bits of the key (security test) as described above to be certain that it is secure. Note that Alice should prepare more qubits, since part of the key is used to test eavesdropping and cannot be used as part of the key that encrypts the message.

\textbf{2. The Eavesdroppers keep Alice's qubit.} This strategy is slightly different from the previous one. Instead of measuring Alice's qubit first, the Eavesdroppers take the qubit and store it without measuring it. They replace Alice's qubit with another one, prepared in a random state, so that Bob can still measure one qubit. In our game, the Eavesdroppers take the qubit student out of the Bloch circle and replace them with another qubit student prepared in one of the states $\{ \ket{0}, \ket{1}, \ket{0_x}, \ket{1_x} \}$ at random (see Figure~\ref{students pos bb84}). The original qubit prepared by Alice is guarded by one of the Eavesdroppers. Once Alice announces the list of preparation axes, the Eavesdroppers measure the guarded qubit (see Section~\ref{rules one qubit}) in the same direction as Alice's preparation axis to always obtain exactly the same bits of the key that Alice has. However, since Bob gets a random state, he obtains the same outcome as Alice only with probability 1/2, even when he measures in Alice's directions. Therefore, the Eavesdroppers are detected with probability 1/2. As for the previous strategy, Alice and Bob can do the security test to detect Eavesdroppers' action.

\textbf{3. The Eavesdroppers try to decrypt the message by guessing the key.} The Eavesdroppers have access to the encrypted message since it is sent via a public channel (said out loud in our game), so they can try and guess the key to decrypt it. With this simple strategy, students can work with probability theory (e.g. to compute the total probability of correctly guessing the key depending on the number of bits $n$\footnote{In case letters are encrypted as binary numbers as presented here, this probability is $1/2^n$.}).

If Alice and Bob detect the action of the Eavesdroppers, they abort the protocol and do not send any message. The Eavesdroppers win if they are able to decipher the secret message without being noticed. On the other hand, Alice and Bob win if they are able to communicate the message secretly and successfully.

\subsection{E91}\label{E91}
\subsubsection{Theoretical background}\label{theory e91}
\begin{figure}[ht!]
\includegraphics[width=3.4in]{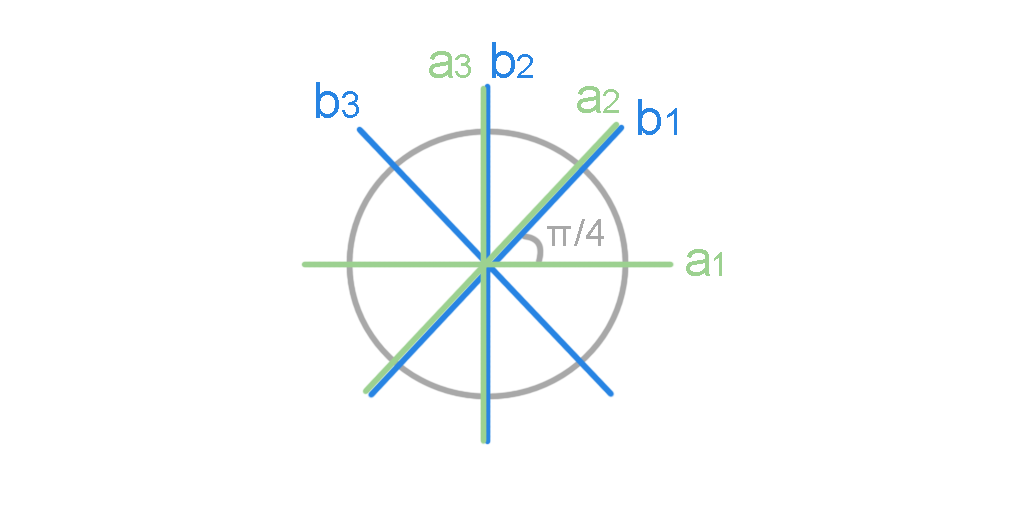}
\caption{Measurement directions for the E91 protocol. The possible directions that Alice and Bob can choose are denoted as $\{a_1, a_2, a_3\}$ and $\{b_1, b_2, b_3\}$ respectively. Note that Alice and Bob both measure in the same direction if they choose, respectively, $(a_3,b_2)$ or $(a_2,b_1)$. In order to perform a Bell test, they measure in the directions depicted in bold.}\label{meas dir e91}
\end{figure}

The goal in this case is the same as for the BB84 protocol, i.e., to generate a secure encryption key. However, the E91 protocol developed by Artur K. Ekert in 1991 \cite{PhysRevLett.67.661} proposes a different procedure to generate a secure key. In this case, an entangled pair of qubits is sent from a source S to two receivers, Alice and Bob. Alice and Bob choose their measurement basis randomly and independently among the ones shown in Figure~\ref{meas dir e91} and announce their chosen orientation via a public channel. The measurements are divided into two groups and there are two possibilities of outcomes: (i) Alice and Bob have chosen the same measurement basis, i.e., due to the properties of entangled qubits, every time Alice gets an outcome 0 (or 1), Bob gets the complementary outcome 1 (or 0). Alice and Bob use these bits for the key. (ii) Alice and Bob have chosen different measurement basis and they need to compute the quantity $S$ of the CHSH inequality with their data to check whether the correlation of the outcomes of the entangled pair of qubits has been classical ($S \leq 2$ and thus CHSH inequality fulfilled) or quantum mechanical ($S > 2$ and thus violates the CHSH inequality). If the CHSH inequality is fulfilled, the communication has been eavesdropped by a third party and thus Alice and Bob have to create a new secure key.

\subsubsection{Rules of the game}
In this case, we also need students playing the roles of Alice, Bob, Eavesdroppers and qubits. The qubits start as entangled pairs (see Figure~\ref{entangled pair} (Up left)).

The goal of this activity is exactly the same as in Section~\ref{BB84}, i.e., sending a secret message successfully. The only difference is how Alice and Bob generate a secure key. Therefore, we refer the reader to Section~\ref{send secret message} for more information on how to encode and decode the message with a key.

In this case, Alice and Bob share a pair of entangled qubits in the state $\ket{\psi^-}$ and they are going to measure them in different directions. In particular, they perform a Bell test to check if the transmission of the key is really secure, so among their measurement directions, some of them should be separated by $\pi/4$ (see Section~\ref{rules CHSH}). All possible measurement directions (Figure~\ref{meas dir e91}) are painted on the Bloch plane on the floor and the positive side of each axis is marked with the compass stick by Alice and Bob at the beginning of the game (they have to agree on this so that the positive side of each axis is the same for both).

As explained in Section~\ref{theory e91}, in order to create a secure key, Alice and Bob measure several pairs of entangled qubits in the state $\ket{\psi^-}$. For each pair, Alice chooses to measure her qubit in one out of the three directions $\{a_1,a_2,a_3\}$ (see Figure~\ref{meas dir e91}). She writes down both the measurement direction she chose and the outcome she obtained. We refer the reader to Section~\ref{rules CHSH} --Figure~\ref{Alice meas CHSH} in particular-- for details on the rules of Alice's measurement process. Then Bob does the same, i.e. he randomly chooses a measurement direction among $\{b_1,b_2,b_3\}$ and writes down his axis choice and the outcome. If the chosen direction was the same as Alice's, the detailed procedure for our game can be seen in Figure~\ref{entangled pair}. Otherwise, the measurement axes are separated by $\pi/4$, i.e. the situation is the same as for a Bell test (see Section~\ref{rules CHSH} and Figure~\ref{Bob meas CHSH}).

After both of them have performed all the measurements, they publicly compare (say out loud) the list of measurement directions. In order to create the key bit string, Alice and Bob take the outcomes of coincident measurement axis. They have to agree on how to transform the outcomes $+1,-1$ into $0,1$. Since they always get anti-correlated results when both measure the state $\ket{\psi^-}$ in the same direction (see Section~\ref{entangled qubits}), the easiest way is: when Alice gets outcome $+1$ ($-1$), she writes the bit $0$ ($1$). Bob does the opposite, i.e., when he gets $+1$ ($-1$), he writes the bit $1$ ($0$). Hence, both get the same bit of the key when they measure in the same direction.

But, how can the Eavesdroppers act in this case? Like in the BB84 protocol, the Eavesdroppers can measure (hit with the ball) Bob's qubit before him and then send the qubit to him, hoping that he would not realize that his qubit has already been measured. However, the Eavesdroppers' measurement broke the initial entanglement, so the qubit Bob receives is no longer entangled. Thus, it is easy for Alice and Bob to detect eavesdropping by just performing a Bell test and checking the CHSH inequality (see Eq.~\eqref{eqCHSH}). After the measurement process, they just need to take the outcomes of the non-coincident measurement directions, compute $|\langle S \rangle |$\footnote{The CHSH inequality reads $|\langle a_1b_1\rangle - \langle a_1b_3\rangle + \langle a_3b_1\rangle  + \langle a_3b_3\rangle | \leq 2$ in this particular case.} and check if it is larger than 2, in which case they know that their qubits were perfectly entangled and there was no Eavesdropper. Otherwise, they discard the key. Once they have the secure key, Alice sends (say out loud) an encrypted message and Bob can decrypt it. Alice and Bob win if they are able to send the secret message successfully without the Eavesdroppers getting access to it, so they have to make sure that their key is secure. The goal of the Eavesdroppers is to intercept the message.

This whole protocol is based on Alice and Bob sharing pairs of entangled particles. Real entangled particles can be separated by an arbitrarily large distance and they are still entangled, which makes the protocol even more useful, since the key can be generated among parties that are far apart from each other. If the teacher wants to emphasize this property, the variant of the game where entangled qubits are placed in two different Bloch circles can be played (see Section~\ref{rules CHSH}).

\section{Summary}
We have introduced a game to teach advanced concepts of quantum mechanics such as Bell tests and quantum cryptography. Students play the role of both scientists and qubits so they can make experiments and analyze the outcomes in the same way as if they were in a real laboratory, getting even the same results! This kinesthetic approach allows the students not only to experience first-hand the properties of quantum mechanics, but also to learn how it is to be a scientist and to develop skills such as critical thinking and creativity.

The Bloch sphere is represented as a circle on the floor (Bloch plane) and the students themselves represent Bloch vectors by orienting their bodies in the corresponding directions. Pairs of entangled qubits are played by two students holding hands and spinning around the center of the circle. The students playing the scientists can measure single and entangled qubits in \textit{any} direction, which allows them to perform advanced experiments of quantum mechanics such as Bell tests. The results the students obtain are the same as if they were working with real qubits, since we use different types of dice to reproduce the stochastic nature of quantum measurements.

In addition, we show how to apply our game to reproduce cryptography protocols such as the BB84 or the E91. By reproducing single-particle states and measurements in any possible direction, we enable the students to perform the BB84 protocol, whereas for the E91 they need to be able to do Bell tests on entangled pair of particles, which they can also reproduce within the framework of our game.

\newpage
\renewcommand\bibname{References}
\bibliography{QuantumCryptography}
\newpage

\appendix
\numberwithin{equation}{section} 
\section{Details on Theoretical background}
\setcounter{equation}{0}
\renewcommand{\theequation}{A.\arabic{equation}}

\subsection{CHSH Inequality}\label{app CHSH}

The CHSH inequality was first mentioned by Clauser et al. in 1969 \cite{PhysRevLett.23.880} and gives constraints on theories which are based on so called local hidden variables \cite{RevModPhys.38.447} and thus are subject to local realism. These local hidden variables arised through the assumption of Einstein, Podolsky and Rosen that quantum mechanics may not be a complete theory and thus become complete by the introduction of these local hidden variables. The CHSH inequality generalizes Bell's theorem \cite{RevModPhys.38.447, PhysicsPhysiqueFizika.1.195} to experimental realizations \cite{PhysRevD.10.526, PhysRevLett.28.938} and reads in its general form
\begin{align}
S= |E(a,b)-E(a,b')+E(a',b)+E(a',b')|, \; S \le 2
\label{eq:Appendix:CHSH}
\end{align}
where the correlation $E(a,b)$ is defined to be the expectation value of the product of the outcomes $A(a)$ and $B(b)$ at measurement settings $a$ and $b$, i.e., the statistical average of $A(a) \cdot B(b)$. The mathematical formulation of the theory of quantum mechanics predicts a maximum value for the quantity $S$ of $S_{\mathrm{QM}}=2\sqrt{2}\approx 2.8284$ which is greater than 2 that one would maximally obtain for theories based on locality and realism. A violation of \eqref{eq:Appendix:CHSH} thus shows that this tested theory, in particular quantum mechanics, does \textit{not} follow the rules of local realism.

This maximal violation of the CHSH inequality due to the unique features of quantum mechanics can be achieved by considering one of the maximally entangled Bell states, e.g. $\ket{\psi^-}$, as used for our approach in the main text.
For this calculation, we first define an operator \mbox{$\hat{A}(a)=\vec{r}(a) \vec{\hat{\sigma}}=\sin(a)\hat{\sigma}^x+\cos(a)\hat{\sigma}^z$} on the Blochsphere at angle $\phi=0$, i.e., our vector $\vec{r}(a)$ rotates in the $x-z$ plane.\footnote{Here, the variable $a$ denotes the angle $\theta_a$ in the $x-z$ plane.} This quantum mechanical expectation value $E(a,b)$ for two configurations $a$ and $b$ in the state $\ket{\psi^-}$ reads
\begin{equation}
E(a,b)=\langle \psi^-| \hat{A}(a) \otimes \hat{B}(b)|\psi^- \rangle=-\textrm{cos}(a-b)
\end{equation}
where $\otimes$ denotes the tensor product.\\
For the specific set $a=\pi/2$, $a'=0$, $b=\pi/4$ and $b'=-\pi/4$ of measurement angles, these quantum mechanical expectation values read $E(a,b)=E(a',b)=E(a',b')=-E(a,b')=-\cos(\pi/4)=-1/\sqrt{2}$. Plugging these values into Eq.~\eqref{eq:Appendix:CHSH}, the quantum mechanical quantity $S_{\mathrm{QM}}$ reads $S_{\mathrm{QM}}=|-4 \cdot 1/\sqrt{2}|=2 \sqrt{2} \approx 2.8284$ which gives the maximal violation of the CHSH inequality by considering the unique features of quantum mechanics.

\subsubsection*{Bell's derivation}
Although Clauser et al. first proposed this inequality, Bell \cite{Bell} gave a general derivation of this formula in 1971 which will be described here.

We first start by defining the measurement outcomes of an operator in certain directions $\vec{a}$ and $\vec{b}$, where \mbox{$\vec{a}=a \cdot \vec{\sigma_1}$} and $\vec{b}=b \cdot \vec{\sigma_2}$ with $\vec{\sigma_1}$ and $\vec{\sigma_2}$ being normalized vectors, can take the values $A(a)=\pm 1$ and $B(b)=\pm 1$. Now we introduce a single continuous local hidden variable, called $\lambda$. In general $\vec{\lambda}$ could be a multi-component vector; yet for the simplicity of the derivation, we will omit the vector properties. Bell thus -- in the spirit of the EPR paradox -- introduced this originally proposed local hidden variable that would make the theory of quantum mechanics complete. The measurement outcomes of two quantities $A$ and $B$ that now depend on both, the measurement direction and the additional local hidden variable, read
\begin{equation}
A(a,\lambda)=\pm 1, \:\:\: B(b,\lambda)=\pm 1.
\end{equation}
Furthermore, we assume locality, i.e., measurement of $A$ is independent on the setting $b$ and measurement of $B$ is independent on $a$, respectively.\\
Let $p(\lambda)$ be the normalized probability distribution of $\lambda$, i.e. $\int_{\mathbb{R}} p(\lambda)d\lambda=1$. Since a source emits entangled particles in two distant directions in a physical manner independent on the parameters $a$ and $b$, $p(\lambda)$ is not dependent on $a$ and $b$ either.

According to Ref.~\cite{RevModPhys.38.447}, we define the expectation value of the inner product of $A$ and $B$ as the correlation function
\begin{equation}
E(a,b)=\int_\Gamma A(a,\lambda)B(b,\lambda)p(\lambda)\:d\lambda
\end{equation}
where $\Gamma$ is the set of all possible values of $\lambda$.\\
With $a'$ and $b'$ being alternative measurement settings, we obtain
\begin{align}
&E(a,b)-E(a,b') \nonumber \\
&=\int_\Gamma \left(A(a,\lambda)B(b,\lambda)-A(a,\lambda)B(b',\lambda)\right)p(\lambda)\: d\lambda \nonumber \\
&=\int_\Gamma A(a,\lambda)B(b,\lambda)[1 \pm A(a',\lambda)B(b',\lambda)]p(\lambda)\: d\lambda \nonumber \\
&-\int_\Gamma A(a,\lambda)B(b',\lambda)[1 \pm A(a',\lambda)B(b,\lambda)]p(\lambda)\: d\lambda.
\label{eq:A.5}
\end{align}
Now, we take the absolute value of \eqref{eq:A.5} and by using the fact that $|A|,|B| \le 1$, we further obtain
\begin{align}
&|E(a,b)-E(a,b')| \le \nonumber \\
&=\int_\Gamma |1 \pm A(a',\lambda)B(b',\lambda)|p(\lambda)\: d\lambda \nonumber \\
&+\int_\Gamma |1 \pm A(a',\lambda)B(b,\lambda)|p(\lambda)\: d\lambda \nonumber \\
&= 2 \pm \int_\Gamma |A(a',\lambda)B(b',\lambda)+A(a',\lambda)B(b,\lambda)| p(\lambda)\: d\lambda \nonumber \\
&= 2 \pm |E(a',b')+E(a',b)|.
\label{eq:A.7}
\end{align}
We can rewrite \eqref{eq:A.7}, by using symmetry properties of the absolute value and the triangle inequality, as
\begin{align}
&|E(a,b)-E(a,b')+E(a',b)+E(a',b')| < \nonumber \\
&|E(a,b)-E(a,b')|+|E(a',b)+E(a',b')|\le 2
\end{align}
which describes the CHSH inequality of Eq.~\eqref{eqCHSH} in the main text.

\setcounter{equation}{0}
\renewcommand{\theequation}{B.\arabic{equation}}
\section{Mermin's approach}\label{Mermin}
In this section, we present an alternative approach to test local realism based on the work "Is the moon there when nobody looks? Reality and the Quantum theory", by N. David Mermin \cite{Mermin}. The main idea is the same, i.e., the quantum properties of the entangled state $\ket{\psi^-}$ contradict the classical concepts of locality and realism. However, in this case there is not a Bell inequality as the CHSH inequality to check. Instead, a different experimental setup is proposed.

There are three different measurement directions separated by an angle of $2\pi/3$ instead of $\pi/4$ as in the CHSH approach highlighted in Section~\ref{sectionCHSH}. As before, if Alice and Bob both measure the state $\ket{\psi^-}$ in the same direction, they always obtain anti-correlated results. However, if they measure in directions separated by $2\pi/3$, the probability of getting anti-correlated results is $25\%$.\footnote{These probabilities are obtained by using the projection operator $\textbf{P}(n,\pm)=\frac{1}{2}(\mathbb{1}\pm \vec{n}\cdot \vec{\sigma})$ onto the states $\ket{0_n}$ ($+$) and $\ket{1_n}$ ($-$) of a generic direction $\vec{n}$. $\vec{\sigma}$ denotes the vector $(\sigma_x,\sigma_y,\sigma_z)$, where $\sigma_{x,y,z}$ are the Pauli matrices. If one denotes Alice's measurement direction as $a$ and Bob's as $b$, the resulting probabilities are:
\begin{eqnarray*}
p\,(+_A,+_B)&=& \bra{\psi^-}\textbf{P}_A(a,+)\textbf{P}_B(b,+)\ket{\psi^-}=\frac{1}{4}(1-\cos \alpha),\\
p\,(-_A,-_B)&=& \bra{\psi^-}\textbf{P}_A(a,-)\textbf{P}_B(b,-)\ket{\psi^-}=\frac{1}{4}(1-\cos \alpha),\\
p\,(+_A,-_B)&=& \bra{\psi^-}\textbf{P}_A(a,+)\textbf{P}_B(b,-)\ket{\psi^-}=\frac{1}{4}(1+\cos \alpha),\\
p\,(-_A,+_B)&=& \bra{\psi^-}\textbf{P}_A(a,-)\textbf{P}_B(b,+)\ket{\psi^-}=\frac{1}{4}(1+\cos \alpha),
\end{eqnarray*}
where $p\,(+_A,+_B)$ is the probability that Alice gets the outcome $+1$ and Bob gets $-1$, etc. and $\alpha$ is the angle between the measurement directions $a$ and $b$. Thus, the probability of getting anti-correlated results when $\alpha=2\pi/3$ is $p_{\text{diff}}=p\,(+_A,-_B)+p\,(-_A,+_B)=1/4$. Note that for the same measurement direction ($\alpha=0$): $p_{\text{diff}}=1$.}

If one considers a data set containing the measurement results of Alice and Bob measuring the state $\ket{\psi^-}$ in all possible combinations of measurement directions, the probability to find two outcomes that are different is
\begin{eqnarray}
p_{\text{diff}}&=&p_{\text{same\,dir}}\cdot p_{\text{diff}}+ p_{\text{diff\,dir}}\cdot p_{\text{diff}}\\
&=&\frac{1}{3}\cdot 1+\frac{2}{3}\cdot \frac{1}{4}=\frac{1}{2},
\end{eqnarray}
i.e., the probability that both measure in the same direction and obtain different results or they measure in different directions and obtain different results.

Let us now consider the case in which we assume realism in our theory, i.e., the properties of the state are fixed before the measurement. Since the results from Alice and Bob by measurement in the same direction are anti-correlated, one can conceive two possible situations in which the qubits have fixed properties and give such results:
\begin{enumerate}
\item Alice's qubit has fixed properties such that it always gives $+1$ when measured in direction $a_1$, $+1$ when measured in $a_2$ and $+1$ when measured in $a_3$. Thus, Bob's qubit should have fixed properties such that it gives $(-1,-1,-1)$ when measured in directions $(b_1,\,b_2,\,b_3)$, respectively, in order for the outcomes to be anti-correlated. In this case, Alice and Bob always get different outcomes no matter their choice of measurement direction.
\item Alice's qubit has fixed properties such that it always gives $+1$ when measured in direction $a_1$, $-1$ when measured in $a_2$ and $+1$ when measured in $a_3$; so Bob's qubit should give $(-1,+1,-1)$ after measurement in $(b_1,b_2,b_3)$, respectively. Thus, the outcomes are different in the cases where Alice and Bob measure in the following directions: $a_1-b_1$, $a_1-b_3$, $a_2-b_2$, $a_3-b_1$ and $a_3-b_3$, i.e. in $5/9$ of the times. All other possible sets of fixed outcomes are analogous to this case and also give different outcomes $5/9$ of the time.
\end{enumerate}

Thus, in order to check their theory for local realism, Alice and Bob just need to make measurements of the state $\ket{\psi^-}$ in all possible combinations of measurement directions and then count the number of times they obtain different outcomes. If they obtain different outcomes $5/9$ of the time or more local realism is still valid, whereas if they obtain different outcomes $1/2$ of the time, then local realism cannot be assumed. In order to distinguish between the probabilities $p_\mathrm{cl}=1/2=0.5$ and $p_\mathrm{qm}=5/9 \approx 0.556$, there should be enough measurements to get good statistics and a reliable result. An estimation of the necessary number of students and measurements per student is given in Appendix~\ref{app statist}.

Within the framework of our proposal, the measurement process of the entangled pair of qubits is exactly the same as the one detailed in Section~\ref{sectionCHSH} in the main text. The only difference in this second approach is that the measurement directions are separated by an angle of $2\pi/3$ instead of $\pi/4$. Alice measures first in the exact same way as described in Figure~\ref{Alice meas CHSH}. Then, if Bob measures in directions that are separated by $2\pi/3$ from Alice's direction, the corresponding qubit should throw the biased die. In this case, the die should be designed such that with a probability of $75\%$ the qubit rotates to the closer side of the axis and with $25\%$ it rotates to the further side (see Figure~\ref{Bob meas CHSH}).

The advantage of this approach is that there is no need to introduce any mathematical concept such as the CHSH inequality in advance. Therefore, we suggest that the activity is conducted as follows: first, the concept of local realism is presented and the students should discuss their own intuitions and ideas with respect to it. Then, they compute the probabilities for the different cases based on local realism hypothesis (see points 1 and 2 in the enumeration above). Finally, the students perform the measurements with the rules of the game we propose and analyze the probabilities they obtain to compare and contrast their initial intuitions with the "real" experiment.

\subsection*{Comparison of approaches}\label{comparison}
We have proposed two different ways of testing local realism to give the teachers the opportunity to choose how the content is presented. On the one hand, the CHSH approach has the advantage that fewer measurements are needed to get good statistics and compute the quantity $S$ (see Appendix \ref{app statist}). However, the CHSH inequality is harder to explain to students without going into much mathematical depth. On the other hand, Mermin's approach provides the possibility to directly compare the initial intuition of the students, presumably in favor of local realism, to the quantum non-intuitive results. Thus, they can experience the whole process of testing an hypothesis (local realism) with experimental results and elaborating explanations after the contradictory results. In this case, the activity should be presented without previous lectures on local realism and the difference with the quantum case, so that the students can experience it by themselves. The downside of this latter approach is that the number of measurements needed to get good statistics is much larger than for the CHSH approach (see Appendix \ref{app statist}).

\section{Statistics}\label{app statist}
\setcounter{equation}{0}
\renewcommand{\theequation}{C.\arabic{equation}}
\setcounter{table}{0}
\renewcommand{\thetable}{C.\arabic{table}}

\subsection{Examples of biased dice}\label{app biased dice}
Here, we depict for some specific sets of measurement settings $a$, $b$, $a'$ and $b'$ examples of the biased die that the students should throw when perfoming the game for the different approaches.

For the Mermin's approach of Appendix \ref{Mermin}, an angle of $2\pi/3$ between each measurement setting of $a$ and $b$ gives with Eqs.~\eqref{prob theta1} and \eqref{prob theta2} from the main text the probabilities $p_0=\cos^2(\pi/3)=1/4$ and $p_1=\sin^2(\pi/3)=3/4$. Thus, the students can use a tetrahedron to perform the game as proposed. A tetrahedron can also be easily built by the students beforehand as an external activity.

For the CHSH inequality of Section~\ref{sectionCHSH} in the main text, an angle of $\pi/4$ between each measurement setting is used. Here, the corresponding probabilities read $p_0=\cos^2(\pi/8) \approx 0.854$ and $p_1=\sin^2(\pi/8) \approx 0.146$. In this case, a possible biased die could be an icosahedron, i.e. 20-sided die. Here, 17 of 20 sides correspond to probability $p_0$, 3 of 20 sides correspond to $p_1$.
The students should build the dice first by themselves as an external activity.
Since the probabilities are not precisely 0.85 and 0.15, respectively, there will be a statistical relative error of $(0.854-0.850)/0.850 \approx 0.5\%$ that cummulates with the number of dice thrown.\\
To circumvent (or at least shrink) this statistical error, freely available apps for smartphones or computers may be used as an alternative multi-faced die (for example, RNG Plus for Android or Roll Dice Online) where one can define the number of sides of the die. For the case described above, for example a 1000-sided die may be thrown. If the number of the simulated thrown die is between 1 and 854, the event with probability $p_0$ takes place; for a number between 855 and 1000, the event with probability $p_1$ takes place.

For the last case of Section~\ref{rules one qubit} in the main text where the $x$ and $z$ directions are the measuring basis, i.e. an angle of $\pi/2$ between each measurement setting, the probabilities are $p_0=p_1=0.5$ which can be realised by throwing a standard die (where e.g. odd numbers correspond to $p_0$ and even numbers to $p_1$) or a standard coin.

An overview over the proposed dice in the different approaches can be seen in Table \ref{Table:dices}.
\begin{table}
\begin{center}
\begin{tabular}{p{1.5cm}|p{1.5cm}p{1.5cm}p{1.5cm}p{2cm}lllll}
Game & angle & $p_0$ & $p_1$ & die \\[0.1cm]
\hline
\rule{0pt}{4ex}
Mermin & $2\pi/3$ & 3/4 & 1/4 & tetrahedron \\
\rule{0pt}{4ex}
CHSH & $\pi/4$ & $\approx$ 0.854 & $\approx$ 0.146  & App, icosahedron\\
\rule{0pt}{4ex}
$x-z$ basis & $\pi/2$ & 1/2 & 1/2  & standard die/coin \\
\hline
\end{tabular}
\end{center}
\caption{Overview over the proposed dice in the different game proposals. The second column (angle) describes the angle between each measurement setting $a, b, a'$ and $b'$.}
\label{Table:dices}
\end{table}

\subsection{Statistical error}\label{app stat error}
The two approaches introduced in the main text require multiple measurements of each quantity whose statistics we describe here. We present some estimations of the number of measurements of the quantity $S$ in the CHSH approach and probability $p$ in the Mermin's approach, respectively, such that the students obtain a statistical reliable answer to the question whether the rules of the game were based on the theory of quantum mechanics or local realism.

As a measure of distinguishability of the students' measurement outcome between quantum mechanics and local realism, we consider the statistical standard deviation of each quantity.
The standard deviation of a quantity which is measured multiple times is defined as
\begin{equation}
\sigma=\dfrac{1}{\sqrt{N}}
\label{sigma}
\end{equation}
where $N$ is the total number of measurements.

For the CHSH approach presented in Section~\ref{sectionCHSH} in the main text, the goal of the proposed game is that the students shall violate the CHSH inequality, i.e., obtaining a value for the quantity $S$ that is larger than 2. This thus tells that the game the students played -- and in particular the theory of quantum mechanics -- does not follow local realism.

We say that we accept the outcome (either CHSH inequality is violated or is not violated) when we can statistically claim the correctness and reliability of the value of our quantity $S$, i.e., we want to statistically distinguish the two outcomes with a high certainty. Here, two errors can occur: Either our null hypothesis $H_0$ (here, quantum mechanics does not follow the theory of local realism) is correct but we erroneously refuse it (type one error) due to our statistics, or our null hypothesis is wrong but we erroneously accept it (type two error). For our significance level $\alpha$ as the probability of error given for the null hypothesis $H_0$, we allow a value of $5\%$, i.e., our confidence of the correct result is $95\%$. In general, the type one error is the graver error which we thus want to avoid.

In the game based on the Mermin's approach of Appendix~\ref{Mermin}, the students have to distinguish between the probabilities $p_\mathrm{qm}=1/2=0.5$ for a game that is based on the features of quantum mechanics and $p_\mathrm{cl}=5/9 \approx 0.556$ for a theory based on local realism. Our motivation is to distinguish between these two probabilities $p_\mathrm{qm}$ and $p_\mathrm{cl}$. We assume a Gaussian distribution for both probabilities for which we need to calculate the standard deviation $\sigma$ such that they are perfectly distinguishable (not overlapping). For a one-tailed test with Gaussian distribution, $95\%$ of the obtained results lay in the interval $I_\mathrm{qm}=[p_\mathrm{qm};\, p_\mathrm{qm}+1.64\sigma]$. Therefore, we set our upper border $I_\mathrm{r}=5/9$ such that the obtained result was due to a game based on the unique features of quantum mechanics with $95\%$ success probability and consequently with significance level $\alpha=0.05$, i.e., $|0.5-5/9|=1.64\sigma$ and thus the standard deviation reads $\sigma \approx 0.034$. Subsequently and according to Eq.~\eqref{sigma}, the total number of measurements to be taken by the students is $N \approx 866$. Alternatively, one can calculate this number for $N$ by making the ansatz that the right border of the interval $I_\mathrm{qm}$ just equals the left border of the interval $ I_\mathrm{qm}=[p_\mathrm{cl}-1.64\sigma; p_\mathrm{cl}]$, i.e. $p_\mathrm{qm}+1.64\sigma = p_\mathrm{cl}-1.64\sigma$ and solve for $N$ and which leads to the same result.

For the CHSH approach of Section~\ref{sectionCHSH} in the main text, the calculation of the statistics is similar. In the CHSH game \cite{CHSHgame}, the classical win probability $p_\mathrm{cl}$ for Alice and Bob to answer at least one of two questions asked by a neutral referee correctly is $p_\mathrm{cl}=0.75$. On the other hand, when Alice and Bob share an entangled pair of qubits, the winning percentage rises up to a maximum value of $p_\mathrm{qm}=0.5+0.5/\sqrt{2} \approx 0.854$. Again we assume a Gaussian distribution of the measured values for the win probability $p$ around its mean value $p_\mathrm{qm} \approx 0.854$ as the maximal winning percentage (thus we have a one-tailed hypothesis test where $95\%$ of the obtained results lay in the interval $I=[p_\mathrm{qm}-1.64\sigma; p_\mathrm{qm}]$. Therefore, $|0.854-0.75|=1.64\sigma$ gives the standard deviation $\sigma \approx 0.063$ and thus $N \approx 250$ measurements to be taken by the students. If we assume around 8 groups of each 3 scientists in the class, each group has to do roughly 30 measurements. As the students playing the role of the qubits and scientists can swap their roles, each student should play the role of the qubit and scientist, respectively, around 15 times.

Here, we point out that all measurement results obtained by the students should be put together to do the statistics and to analyse the obtained data, because the average of the averages of a quantity is in general not equivalent to the average of the whole measurement results for a quantity.
\end{document}